# Optimal matching of thermal vibrations into carbon nanotubes

K.G.S.H. Gunawardana and Kieran Mullen
Homer L. Dodge Department of Physics and Astronomy
The University of Oklahoma
440 W Brooks St., Norman,OK,73019

## ABSTRACT

Carbon nanotubes (CNTs) are promising candidates to improve the thermal conductivity of nano-composites. The main obstacle to these applications is the extremely high thermal boundary (Kapitza) resistance between the CNTs and their matrix. In this theoretical work our goal is to maximize the heat flux through the CNT by functionalizing the CNT ends. We use a Landauer approach to calculate and optimize the energy flux from a soft to a hard material in one dimension through a connecting continuous medium of varying elasticity and density. The transmission probability of phonons through the system is calculated both numerically and analytically. We find that over 90% of the maximum heat flux into CNT is possible for 1nm length of the intermediate material at room temperature (300K).

## 1. INTRODUCTION

Thermally conducting polymers have drawn as much attention in recent years as their electronic and ionic counterparts. This new class of materials will have a vast range of applications such as in air craft to replace heavy metallic parts, in defense systems to construct thermal shields and in computers to obtain directional heat flows that can be used in future micro-electronic devices for thermal management. In achieving these goals composite materials of polymer and highly thermal conducting nano-scale inclusions are of great interest [1-3]. These composite materials integrate the pristine properties of polymers with the enhanced thermal conductivity of the inclusions. Carbon nanotubes (CNTs) are promising candidates to improve the thermal conductivity of  nano-composites. The thermal conductivity of SWNTs at room temperature is about 6000W/mK and that of MWNTs are about 3000 W/mK [4-6]. But the realization of this hybrid composite has been greatly hindered by the interface thermal resistance (Kapitza thermal resistance)[7-12]. The Kapitza thermal resistance is the resistance to the heat transport across the boundary of two dissimilar materials[13,14]. The effect becomes very significant when the two materials have a huge difference in elasticity so that there is only a weak coupling of phonon modes at the interface[15,16]. The elastic modulus of CNT is about 1 TPa whereas that of polymer is of the order of 1 GPa [17,18]. Therefore the interface of CNTs and polymer pose a considerable Kapitza thermal resistance. The Kapitza thermal resistance is also becoming a key problem in electronic devices as rapidly shrinking their size [19,20].

 At this point it is essential to have a better understanding of interfacial thermal transport and what modification of the interface is required to maximize the thermal conductance. Kosevish has shown theoretically the existence of low frequency resonant vibrational modes at the interface transition layer between soft and rigid materials [21]. These resonant vibrational modes can reduce the Kapitza thermal resistance at elevated temperatures.  Clancy and Gates have reported a significant reduction of the Kapitza thermal resistance of CNT-Polymer composites by grafting hydrocarbon chains to the surface of the CNT with covalent bonds [22]. Their work is based on molecular dynamic simulations. Moreover they observed further reduction of the Kapitza thermal resistance as increasing the length of the hydrocarbon chains and the grafting density. The above approaches have difficulties in optimizing the properties of the interface transition layers or the grafted atomic chains to obtain the maximum heat flux through the interface due to the mathematical complexity and the huge time cost in MD simulations. Our goal is to engineer the interface transition layer to maximize the heat transfer through the interface between hard and soft materials such as CNTs and polymer through a model calculation.

In section 2 below we describe our model system used for the optimization process and in the section 3 we describe the numerical and analytical approaches to optimizing the interface and the results. There, we consider the propagation of longitudinal vibrational waves in a one dimensional inhomogeneous constriction. These inhomogeneities are the variation of the mass density and the elastic modules from soft to hard materials. Finally, in section 4 conclusion and an outline of future work is presented.

## 2. THE MODEL

Our model consists of a one dimensional chain of atoms (channel) strongly bonded to each other, where the force constants are equivalent to that of carbon nanotubes. This atomic chain is connected to two heat baths at temperatures $T_{Hot}$ and $T_{Cold}$ at either ends. These heat baths replicate very soft media like polymers and have comparably small force constants. The ratio of the force constants between the heat bath and the channel is kept at 1:1000. The coupling (interface) of the heat baths to the channel is always through an atomic chain whose atomic mass and the force constants vary with position.

We adopt the Landauer-Buttiker formalism to describe the thermal flux through the channel $\dot{Q}_C$. In this formalism the ballistic transport of phonons between two phonon reservoirs is given by the Landauer formula [23].

$$\dot{Q}_C = \sum_\alpha \int_0^\infty \frac{dk}{2\pi} \hbar \omega_\alpha(k) \, \upsilon_\alpha(k) \, (\eta_{Hot} - \eta_{Cold}) \, \Gamma(\omega) \qquad [1]$$

Where $k$ is the phonon wave vector, $\omega_\alpha(k)$ and $\upsilon_\alpha(k)$ are phonon frequency and the velocity of the mode $\alpha$, $\Gamma(\omega)$ is the transmission probability of a phonon between the heat baths and $\hbar$ is the Planck constant. The functions $\eta_{Hot}$ and $\eta_{Cold}$ are the Bose Einstein distribution functions of heat baths at temperatures $T_{Hot}$ and $T_{Cold}$, $\eta = (Exp[\hbar\omega/k_BT]-1)^{-1}$ and $k_B$ is the Boltzmann constant.

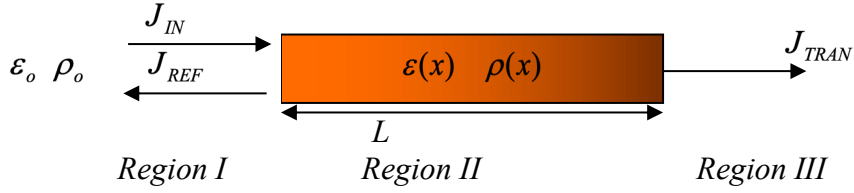

Figure.1 Three regions of interest. Region I is the heat bath, region II is the inhomogeneous interface and the region III is the uniform channel, representing the carbon nanotube.

Ballistic transport of phonons through the channel can be obtained at very low temperatures where the interactions of phonons are negligible. In this limit the transmission probability $\Gamma(\omega)$, is determined by the smoothness of the coupling between the heat baths and the channel. When there is perfect coupling, $\Gamma(\omega) = 1$, and we have the maximum possible heat flux through the channel; $\dot{Q}_{\max} = k_B \pi^2 T \Delta T / 3h$. The quantity $k_B \pi^2 T / 3h$ is called the quantum of thermal conductance [23,24]. In our model we assume the channel is a perfect phonon conductor. This approximation is reasonable for CNTs even at higher temperatures. It has been predicted that the ballistic length of CNT can be larger than 100nm at room temperature [25-27]. However our goal is to optimize the interface hence the interaction within the channel is irrelevant. To estimate the transmission probability we use a continuum approximation in which the model system is considered as a one dimensional rod of continuous material. This approximation has been widely used in determining the transmission probability of low frequency acoustic phonons [23,28,29,30]. This approximation remains valid when the dominant phonon wavelength is larger than the lattice spacing [31]. For CNT the temperature at which the dominant phonon wavelength becomes comparable to the lattice spacing is much higher than 300K.

We are first interested in the transmission probability of phonons into the channel (the CNT) from the heat bath (the polymer) and denote this quantity by $\tau(\varepsilon(x), \rho(x), \omega)$. The method of deriving the transmission probability is described below; it is trivial to extend the procedure to calculate $\Gamma(\varepsilon(x), \rho(x), \omega)$, the transmission probability from the polymer, through the CNT and back into the polymer again. Fig.1 depicts the three regions of interest: region I is the uniform, soft medium, region II is the inhomogeneous interface and region III is the uniform, stiff channel. For the optimization we assume region I is in equilibrium at temperature $T_{Hot}$ and III at $T_{Cold}$ and calculate the energy flux through the interface (region II) using equation (1) with $\Gamma$ replaced by $\tau$. Since the wave vector varies with position in region II we convert the integral over $k$ in equation (1) to one over $\omega$ and note the phonon density of states cancels the phonon velocity. The elastic modulus and the mass density of the heat baths and the channel are denoted by $Y_p$, $D_p$ and $Y_c$, $D_c$. The interface region has a position dependent mass density $D(z)$ and elastic modulus $Y(z)$. Before proceeding, we switch to dimensionless quantities. The elastic modulus and the density are expressed in terms of those of the channel, and the length is scaled by the length of the interface region ($L$). That is $x = z/L$, $\varepsilon(x) = Y(x)/Y_C$, $\rho(x) = D(x)/D_C$, $\varepsilon_o = Y_P/Y_C$ and $\rho_o = D_P/D_C$. At left end of region II (x=0) we have $\varepsilon_0 = 0.001$ and $\rho_o = 0.5$ while at the right (x=1) have $\varepsilon_1 = 1$ and $\rho_1 = 1$. The atomic displacements of these three regions are given by $U_I$, $U_{II}$ and $U_{III}$ and only the propagation of longitudinal vibrational waves is considered. In the continuum limit the displacement waves are governed by the scalar wave equation [32].

$$\frac{\partial}{\partial x}\left(\varepsilon(x)\frac{\partial U(x,t)}{\partial x}\right) = \rho(x)\frac{\partial^2 U(x,t)}{\partial t^2} \quad [2]$$

The solutions to the scalar wave equation in the uniform regions take the usual harmonic form given by:

$$U_I(x,t) = A\,Exp[i(k_0 x - \omega t)] + B\,Exp[-i(k_0 x - \omega t)] \quad [3]$$

$$U_{III}(x,t) = C\,Exp[i(k_1 x - \omega t)] \quad [4]$$

where $k_0$ and $k_1$ are wave vectors of the vibrational waves in the heat bath and the channel, given by $k_i = \omega\sqrt{\rho_i/\varepsilon_i}$. In the region II, it is the solution to equation (2). It is impossible to find an exact solution to equation [2] for arbitrary functions $\varepsilon(x)$ and $\rho(x)$. In section 3 below, we give numerical and approximate analytical solutions to this problem.

## 3. SOLUTIONS OF THE MODEL

### 3.1 Numerical optimization.

For given functions $\varepsilon(x)$ and $\rho(x)$ we can find two linearly independent numerical solutions to equation [3] by choosing solutions to different boundary conditions. Specially, we choose $U_{II}(0)=0$, $U'_{II}(0)=1$ and $U_{II}(0)=1, U'_{II}(0)=0$. We then impose continuity on the displacements $U_I$, $U_{II}$ and $U_{III}$ and their derivatives at the boundaries. From this we can calculate the transmission coefficient $\iota = C/A$. The transmission probability of a phonon can be calculated by considering the energy fluxes associated with the displacement waves. The total mechanical energy density $\xi(x,t)$ of the displacement wave $U(x,t)$ can be expressed as the sum of the kinetic and potential energy densities:

$$\xi(x,t) = \frac{1}{2}\rho(x)\left(\frac{\partial U(x,t)}{\partial t}\right)^2 + \frac{1}{2}\varepsilon(x)\left(\frac{\partial U(x,t)}{\partial x}\right)^2 \quad [5]$$

The continuity equation,

$$\frac{\partial \xi(x,t)}{\partial t} + \nabla \cdot J(x) = 0 \quad [6]$$

enables us to solve for the energy flux J(x).

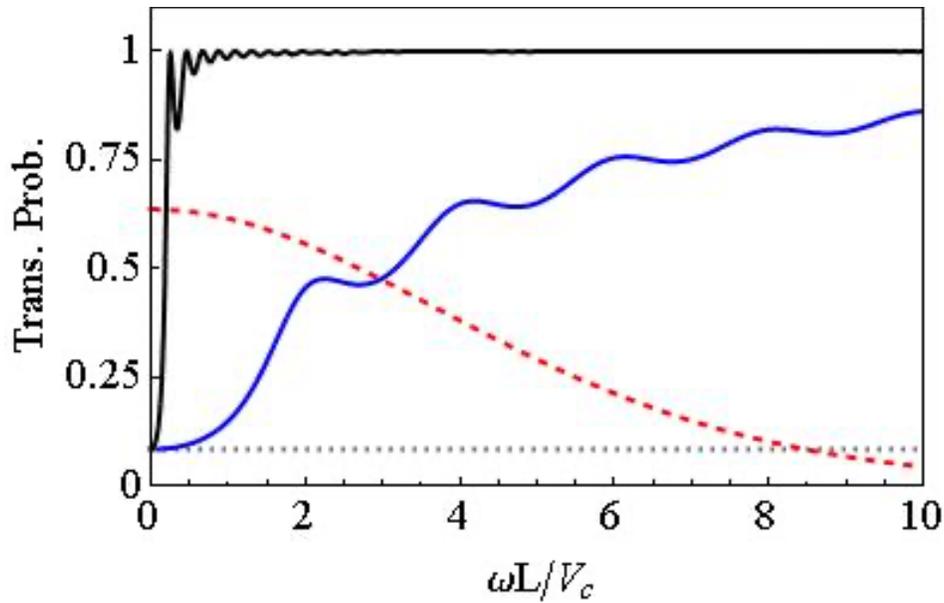

Figure.2 Transmission probability of phonon into the channel from the heat bath vs. scaled frequency. The black dotted line is the transmission probability for an abrupt change of elasticity and mass density and the thick blue line is for a linear interpolation of $\varepsilon(x)$ and $\rho(x)$. The thick black line shows the transmission probability of the optimal variation of $\varepsilon(x)$ and $\rho(x)$. The dashed red line is the thermal weight to the transmission probability which contributed to the Landauer energy flux at $T = 300K$ and $V_c$ is the speed of sound in the channel.

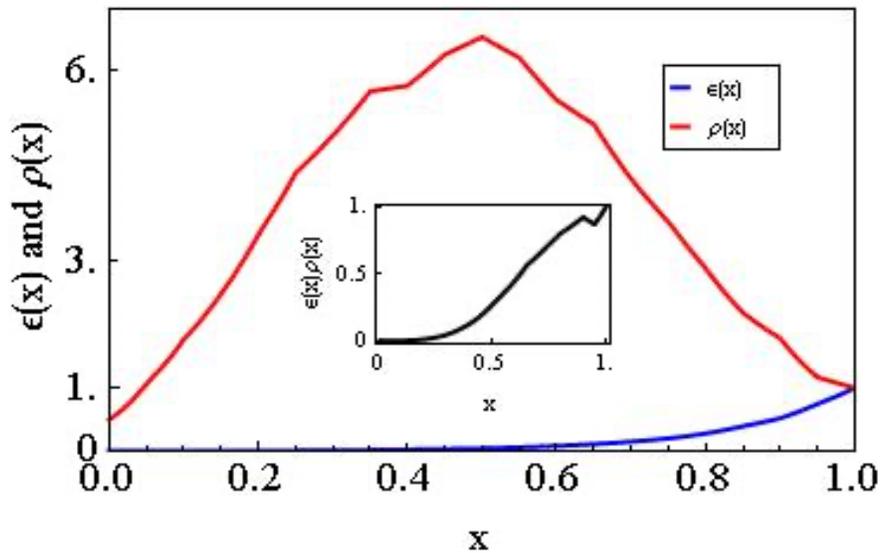

Figure.3 Optimal dimensionless stiffness $\varepsilon(x)$ and density $\rho(x)$ obtained numerically, giving the maximal thermal conductivity. The inset is the product of $\varepsilon(x)$ and $\rho(x)$ vs. position.

For the solutions of the form, $U(x,t) = U(x) e^{i\omega t}$ the mechanical energy flux $J(x)$ can be written as follows.

$$J(x) = i\omega\varepsilon(x)\{U^*(x)\nabla U(x) - U(x)\nabla U^*(x)\} \qquad [7]$$

Then the transmission probability of a phonon into the channel from the heat bath $\tau(\varepsilon(x), \rho(x), \omega)$ can be written as,

$$\tau(\varepsilon(x), \rho(x), \omega) = \frac{J_T}{J_{In}} = \frac{\iota^*(\varepsilon(x), \rho(x), \omega)\, \iota(\varepsilon(x), \rho(x), \omega)}{\sqrt{\varepsilon_0 \rho_0}} \qquad [8]$$

The optimization was started from a linear interpolation of $\varepsilon(x)$ and $\rho(x)$. After calculating the Landauer heat flux through the interface, the functions $\varepsilon(x)$ and $\rho(x)$ were randomly varied subject to the constraints that they take on the correct values at the boundaries and that they are positive. Variations that increased the heat flux were accepted. The process was iterated and the size of the variations decreased until $\varepsilon(x)$ and $\rho(x)$ converged. Fig.2 shows the transmission probability of phonons into the channel from a heat bath against the scaled frequency. The red dashed line is the thermal weight to the transmission probability which contributed to the Landauer energy flux. This is plotted at $T = 300K$ when $\Delta T \ll T$. The black dotted line is the transmission probability for an abrupt change of elasticity and mass density and the blue line is for a linear interpolation of $\varepsilon(x)$, $\varepsilon(x) = (1-\varepsilon_0)x + \varepsilon_0$ and $\rho(x)$, $\rho(x) = (1-\rho_0)x + \rho_0$. The black line shows the transmission probability of the optimal variation of $\varepsilon(x)$ and $\rho(x)$, which is shown in Fig.3. According to the Fig.2 by increasing the length of the interface (L), over which $\varepsilon(x)$ and $\rho(x)$ are varied the transmission function can be squeezed towards low frequencies hence the thermal flux into the channel is increased.

The optimal mass density variation peaks at approximately six times that of the channel density in the middle region of the interface and the elasticity variation is slowly varying on the side of the soft heat bath and has a rapid rise near the channel. This kind of elasticity variation seems physically achievable. There can be a few atomic layers of intermediate force constants decaying rapidly at the interface of hard and soft material. At room temperature (300K) the abrupt change of the material gives only 10% of the maximum heat flux into the channel. With the linear interpolation we can get about 47% of the maximum flux into the channel having 1nm long interface region (L). For the optimal variation 97% of the maximum flux into the channel can be obtained.

It is easy to extend the above techniques to calculate the transmission probability from a soft medium, through one interface, into a stiff channel and then through a second, identical interface and out the other side. Fig.4 shows the variation of energy flux through such a system as a function of the length of the interface region (L) at temperatures 10K, 100K and 300K. In this calculation the energy flux is normalized to the maximum possible energy flux of a phonon mode. As mentioned before by increasing the length of the interface (L), over which $\varepsilon(x)$ and $\rho(x)$ are varied the transmission function can be squeezed towards low frequencies hence the thermal flux into the channel is increased.

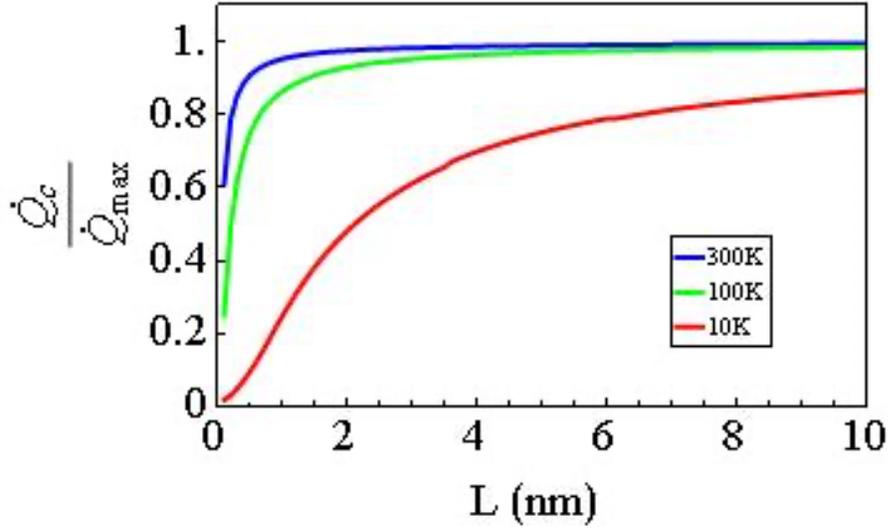

Figure.4 Normalized energy flux through the channel vs. length of the interface at *T=10K, 100K* and *300K*. The channel has an optimized interface at either end connecting it to heat baths at different temperatures.

## 3.2 Analytic approximations.

At discontinuities of material properties, the ratio of the reflected ($\sigma_r$) and incident ($\sigma_i$) stress pulses of longitudinal waves can be expressed as follows [32].

$$\frac{\sigma_r}{\sigma_i} = \frac{\varepsilon_2 \rho_2 - \varepsilon_1 \rho_1}{\left(\sqrt{\varepsilon_2 \rho_2} + \sqrt{\varepsilon_1 \rho_1}\right)^2} \quad [9]$$

Reflectionless transmission can be achieved whenever the numerator of the above equation is zero. That is when $\varepsilon_2 \rho_2 - \varepsilon_1 \rho_1 = 0$. By extending the same argument to a rod whose properties are varying continuously, the condition for reflectionless transmission can be expressed as $\frac{d(\varepsilon(x)\rho(x))}{dx} \approx 0$. This is not always possible when there are given properties at either ends. The inset of Fig.4 shows the product of $\varepsilon(x)$ and $\rho(x)$ against the position for the optimal case. The variation has small derivatives, in agreement with the above condition. The next question raised is whether this optimal variation is unique. We claim that this variation is unique and a reasonable qualitative explanation can be obtained by looking at WKB approximate solution to the equation [3]. The WKB solution to equation [3] can be written as follows:

$$U_{II}(x) = \left(\varepsilon(x)\rho(x)\right)^{-1/4} \exp\left[i\omega \int_0^x \sqrt{\frac{\rho(s)}{\varepsilon(s)}}\, ds\right] \quad [10]$$

The approximation holds when $\frac{d(\varepsilon(x)\rho(x))^{-1/4}}{dy} \ll 1$, Where $y = \int \varepsilon(x)^{-1} dx$ and for higher frequencies $\omega \gg 1$ [33,34]. The transmission probability can then be calculated as in section 3.1 replacing the numerical solution with the WKB solution. The transmission probability obtained numerically and by the WKB method well matches beyond the first resonance peak (Fig.5).

Further understanding of the optimal variation can be obtained analyzing the low frequency ($\omega \ll 1$) behavior of the transmission probability. Low frequency solution to the eq. [3] can be obtained using the perturbation series method [35]. Here we assumed solutions of the form,

$$U_{II}(x) = \sum_{0}^{\infty} f_n(x)\, \omega^{2n} \qquad [11]$$

Then we plug the above solution to the equation [2] and obtain the expression,

$$\varepsilon'(x) \sum_{0} f_n'(x)\, \omega^{2n} + \varepsilon(x) \sum_{0} f_n''(x)\, \omega^{2n} = -\rho(x) \sum_{0} f_n(x)\, \omega^{2(n+1)} \qquad [12]$$

The functions $f_n(x)$ are determined by matching $\omega$ in order by order and with the boundary conditions explained below. Basically we are looking for two linearly independent solutions: $U_{II}^{(a)}$ and $U_{II}^{(b)}$. These two solutions are established by the boundary conditions $U_{II}^{(a)}(0)=1$, $U_{II}^{(a)}(1)=0$ and $U_{II}^{(b)}(0)=0$, $U_{II}^{(b)}(1)=1$. Since the zeroth order term of equation [11] gives the largest contribution the boundary conditions are fully imposed on $f_0(x)$ and all $f_n(0)$ and $f_n(1)$ for $n \geq 1$ are set to zero. The displacement $U_{II}(x)$ was calculated up to the fourth order of $\omega$ and the transmission probability was calculated as in section 3.1 replacing the numerical solutions with the perturbation series solution. The solid brown line in the Fig.5 shows the transmission probability obtained by this method.

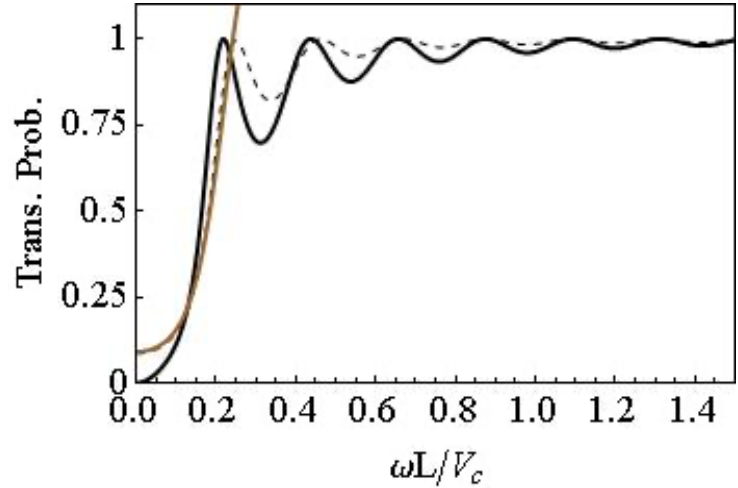

Figure.5 Transmission probability of a phonon into the channel vs. scaled frequency for optimal $\varepsilon(x)$ and $\rho(x)$. The dashed line is from the numerical method. The thick black line is from the WKB approximation and the brown line is from the perturbation series method.

Ideally one would like an analytic form for the optimal stiffness and density variations plotted in Figure 3. While the WKB analysis indicates that it is beneficial to move the first resonance peak lower, it does not give any explicit form for these optimal functions. They may be obtainable from the perturbation series solutions, by inserting the transmission coefficients they generate (symbolically) into equation (1) and then applying the calculus of variations to determine the optimal $\varepsilon(x)$ and $\rho(x)$. This work is underway.

## 4.    CONCLUSION

Thermal transport between soft and stiff media was optimized theoretically. We included an interface variation of material to maximize the thermal transport and the optimal variation of mass density and the elastic modulus of this interface was obtained numerically. We have found the optimal variation can improve the thermal flux to over 90% of its theoretical maximum. Increasing the mass density of the interface can lower characteristic resonances and enhance the low frequency conductivity. It is also found that increasing the length over which $\varepsilon(x)$ and $\rho(x)$ varies improves the thermal conductivity.

The model solved above is simplistic: it ignores nonlinear interactions, it approximates the discrete, atomic system by continuum elasticity theory, and it is one dimensional. However, we feel that the essential insights are sound. That is, including the above physics will not greatly improve the thermal conductivity results, they will just give a more accurate answer showing how poor it is. Similarly, by ignoring the scattering of phonons within the channel (or carbon nanotube) we are overestimating the thermal conductivity. A more detailed model would include a Boltzmann approach to modeling the temperature distribution along the channel. Again, this would not deny the importance of the Kapitza resistance, nor our suggestions for addressing it.

An analytical calculation of the optimal $\varepsilon(x)$ and $\rho(x)$ will be presented in the future. In addition, molecular dynamics simulations will be carried out to compare the optimal variation.